# Robust PI Control Design Using Particle Swarm Optimization

Saeed Tavakoli and Amir Banookh

**Abstract**— This paper presents a set of robust PI tuning formulae for a first order plus dead time process using particle swarm optimization. Also, tuning formulae for an integrating process with dead time, which is a special case of a first order plus dead time process, is given. The design problem considers three essential requirements of control problems, namely load disturbance rejection, setpoint regulation and robustness of closed-loop system against model uncertainties. The primary design goal is to optimize load disturbance rejection. Robustness is guaranteed by requiring that the maximum sensitivity is less than or equal to a specified value. In the first step, PI controller parameters are determined such that the IAE criterion to a load disturbance step is minimized and the robustness constraint on maximum sensitivity is satisfied. Using a structure with two degrees of freedom which introduces an extra parameter, the setpoint weight, good setpoint regulation is achieved in the second step. The main advantage of the proposed method is its simplicity. Once the equivalent first order plus dead time model is determined, the PI parameters are explicitly given by a set of tuning formulae. In order to show the performance and effectiveness of the proposed tuning formulae, they are applied to three simulation examples.

——————————— ◆ ———————————

## 1 INTRODUCTION

Proportional-integral-derivative (PID) control is the most popular controller in the process industry despite the continual advances in control theory. It is a simple and useful controller which gives a powerful solution to the control of a large number of industrial processes. A well designed and adequately tuned PID controller meets most control objectives [1].

Most of PID controllers are proportional-integral (PI) controllers because the derivative action is very often not used. PI control is sufficient for a large number of control processes, particularly when dominant process dynamics are of the first order and the design requirements are not too rigorous [2]. Therefore, good PI tuning methods are extremely desirable due to their widespread use.

Good load disturbance rejection is generally the primary objective. Moreover, the closed-loop system should be robust against model errors. The idea to use a constraint on the maximum sensitivity was proposed in [3]. The use of both maximum sensitivity and maximum complementary sensitivity as design parameters was suggested in [4]. To consider both performance requirements and robustness issues, the design method is aimed at optimizing load disturbance rejection with a constraint on the maximum sensitivity. Good setpoint regulation is also obtained using setpoint weighting. This plays a significant role in improving the setpoint response but has no influence on the load disturbance response.

Approximation of high order processes by low order plus dead time models is a common and well accepted practice. A large number of industrial plants can be approximately modelled by a first order plus dead time (FOPDT) transfer function, as shown in (1).

$$G_p(s) = \frac{K_p e^{-\tau_d s}}{Ts+1} \qquad (1)$$

A FOPDT model does not capture all the features of a high order process, however, it often reasonably describes the process gain, dominant time constant and effective dead time of such a process [5]. Considering the importance of this category of industrial plants, optimal PI tuning formulae for FOPDT processes are proposed in this paper.

## 2 CONTOROL REQUIRMENTS

### 2.1 Load Disturbance Rejection
Load disturbances are the most common disturbances in process control. These low frequency signals are added to the control signal at the process input and drive the system away from its desired operating point [2]. Good rejection of such signals is the first design goal.

### 2.2 Robustenss To Model Uncertaintis
The controller parameters are typically obtained from the model parameters. Because of model uncertainties, the controller parameters should be chosen in such a way that the closed-loop system is not too sensitive to variations in process dynamics. As shown in (2), sensitivity to

————————————————

- *S. Tavakoli is with the Faculty of Electrical and Computer Eng., The University of Sistan and Baluchestan, Iran.*
- *A. Banookhi is with the Faculty of Electrical and Computer Eng., The University of Sistan and Baluchestan, Iran.*

modelling errors is expressed as the largest value of the sensitivity function.

$$M_s = \max_{\omega} \left| \frac{1}{1+G_p(j\omega)G_c(j\omega)} \right| \quad (2)$$

where $M_s$ is the inverse of the shortest distance from the Nyquist curve of the loop transfer function to the critical point. Smaller values of $M_s$ show little or no overshoot while larger ones lead to faster responses. To obtain a robust controller, a constraint on the maximum sensitivity can be used.

## 2.3 Setpoint Regulation

Although the primary design goal is to reject load disturbance signals, it is also important to have good setpoint responses. The main design goal may result in bad setpoint responses because responses to load disturbance and setpoint signals are usually conflicting. As the secondary design goal, good setpoint responses are obtained using setpoint weighting.

## 3 PARTICLE SWARM OPTIMIZATION

In this section, a brief description of particle swarm optimization (PSO) is given. Inspired by social behavior of bird flocking or fish schooling, PSO is a population based stochastic optimization technique developed by Kennedy and Eberhart in 1995.

Having initialized the optimization system with a random population of individuals, optimal solutions are obtained by updating generations. In an n-dimensional search space, the position and velocity of each particle is given by $x_i = [x_{i1}, x_{i2},...,x_{in}]^T$ and $v_i = [v_{i1}, v_{i2},...,v_{in}]^T$, respectively. The position of particle indicates the possible solution in the n-dimensional search space, whereas its velocity indicates the amount of change between the current and next positions.

Corresponding to the personal best solution obtained so far at time t, each particle has its own best position, $p_i = [p_{i1}, p_{i2},...,p_{in}]^T$. The global best particle, $p_g$, represents the best particle found so far at time t in the entire swarm [6], [7]. The new velocity of each particle is calculated by (3).

$$\begin{aligned} v_{ij}(t+1) &= wv_{ij}(t) + c_1 r_1 (p_{ij} - x_{ij}(t)) \\ &+ c_2 r_2 (p_{gj} - x_{ij}(t)), j = 1,2,...,n \end{aligned} \quad (3)$$

The position of each particle is updated in each generation according to (4)

$$x_{ij}(t+1) = x_{ij}(t) + v_{ij}(t+1), j = 1,2,...,n \quad (4)$$

where w is the inertia weight and $c_1$ and $c_2$ are called acceleration coefficients. $r_1$ and $r_2$ are two independent random numbers.

## 4 PROBLEM FORMULATION

Consider a two-degree of freedom structure, as shown in Figure 1

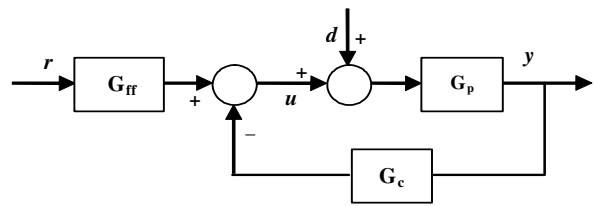

Fig. 1. Block diagram of two-degree of freedom control system.

where r, d and y refer to the setpoint, load disturbance and output signals, respectively. $G_p(s)$ refers to the process model whereas $G_c(s)$ and $G_{ff}(s)$ are PI and feedforward controllers shown in (5) and (6), respectively.

$$G_c(s) = K_c(1 + \frac{1}{T_i s}) \quad (5)$$

$$G_{ff}(s) = K_c(b + \frac{1}{T_i s}) \quad (6)$$

Equation (7) describes the closed-loop control system.

$$Y = \frac{G_p(s)G_{ff}(s)}{1+G_p(s)G_c(s)}R + \frac{G_p(s)}{1+G_p(s)G_c(s)}D \quad (7)$$

The input output relationship for the controller is described by (8).

$$u(t) = K_c \left( br(t) - y(t) + \frac{1}{T_i} \int_0^t (r(\tau) - y(\tau))d\tau \right) \quad (8)$$

The design objective is to determine $G_c(s)$ and $G_{ff}(s)$ to obtain good load disturbance and setpoint responses. A constraint on maximum sensitivity is used to guarantee robustness to model uncertainties. In this paper, $M_s = 1.6$ is considered as the robustness constraint.

## 4.1 Load Disturbance Response

The objective function is to minimize the IAE criterion, shown in (9), subject to a constraint on maximum sensitivity.

$$IAE = \int_0^\infty |r(t) - y(t)| dt \quad (9)$$

The design procedure has two steps. In the first step, the setpoint signal is considered to be zero and $G_c(s)$ is determined so that load disturbances are attenuated and the robustness constraint is satisfied. For the $G_c(s)$ determined in the first step and in absence of load disturbances, $G_{ff}(s)$ is then tuned to achieve good setpoint responses, in the second step.

In order to obtain the optimal PI tuning formulae for the FOPDT model in (1), the PI parameters can be defined based on the model parameters, as shown in (10) and (11).

$$K_c = f_1(K_p, \tau_d, T) \quad (10)$$

$$T_i = f_2(K_p, \tau_d, T) \quad (11)$$

Functions $f_1$ and $f_2$ should be determined such that the load disturbance response is minimized and the robustness constraint is satisfied. However, it is very difficult to determine these functions because each parameter of the controller is a function of three parameters of the model. In order to overcome this difficulty, the procedure for determining $f_1$ and $f_2$ is simplified using dimensional analysis [8].

Considering the process model in (1), the unit of both dead time ($\tau_d$) and time constant ($T$) is the second. The unit of process gain ($K_p$) depends on the nature of the system. Because process gain along with either dead time or time constant cover all the units in (10) and (11), there is only one dimensionless number in the model, namely $\tau_d/T$, which is named dimensionless dead time. Considering the PI controller in (5), the unit of integral time ($T_i$) is the second. The unit of controller gain is the inverse of the unit of process gain. As a result, other dimensionless numbers for the combined model and controller are dimensionless gain ($K_pK_c$) and dimensionless integral time ($T_i/\tau_d$ or $T_i/T$). Based on Buckingham's pi theorem [8], these dimensionless numbers are functions of the dimensionless number in the plant model. Therefore, the PI parameters can be obtained through determining $K_pK_c$ and $T_i/\tau_d$ (or $T_i/T$) from $\tau_d/T$, as shown in (12) and (13).

$$K_pK_c = g_1(\frac{\tau_d}{T}) \qquad (12)$$

$$\frac{T_i}{\tau_d} = g_2(\frac{\tau_d}{T}) \qquad (13)$$

In order to determine $g_1$ and $g_2$ and generate PI tuning formulae, the following procedure is proposed.

**Step 1.** The values of $\tau_d/T$ are selected.

**Step 2.** For each value of $\tau_d/T$, the optimal values of $K_c$ and $T_i$ that minimize the desired objective function are determined using PSO [6], [7].

**Step 3.** The optimal values of $K_pK_c$ and $T_i/\tau_d$ versus $\tau_d/T$ are drawn.

**Step 4.** $g_1$ and $g_2$ are determined using curve fitting techniques.

The values of dimensionless dead time are considered from 0.1 to 2 to consider FOPDT processes with small, medium and fairly long dead time. Table 1 shows the optimal values of $K_pK_c$ and $T_i/\tau_d$, resulting from step 2.

The optimal values of the dimensionless gain and the dimensionless integral time across the selected values of the dimensionless dead time are shown in Figures 2 and 3, respectively. It can be seen from Figure 2 that the dimensionless gain is a function of the dimensional dead time as shown in (14). Similarly, the values of $T_i/\tau_d$ are determined from the values of $\tau_d/T$, using (15).

$$K_pK_c = A_1 + \frac{B_1}{\frac{\tau_d}{T}} \qquad (14)$$

$$\frac{T_i}{\tau_d} = \frac{A_2\frac{\tau_d}{T} + B_2}{\frac{\tau_d}{T} + C_2} \qquad (15)$$

Using the least squares method, $A_1$, $B_1$, $A_2$, $B_2$ and $C_2$ are determined for the best match with Table 1. The optimal values of $A_1$, $B_1$, $A_2$, $B_2$ and $C_2$ are 0.157, 0.415, 0.06, 1.48 and 0.27, respectively.

TABLE 1
OPTIMAL PARAMETERS FOR A FOPDT MODEL

| $\tau_d/T$ | $K_pK_c$ | $T_i/\tau_d$ | b |
|---|---|---|---|
| 0.1 | 4.0088 | 3.7464 | 0.5775 |
| 0.2 | 2.2234 | 3.2147 | 0.8291 |
| 0.3 | 1.5876 | 2.7900 | 0.9359 |
| 0.4 | 1.2512 | 2.4391 | 1.0707 |
| 0.5 | 1.0555 | 2.2178 | 1.1483 |
| 0.6 | 0.8594 | 1.7280 | 1.1588 |
| 0.7 | 0.7527 | 1.5348 | 1.2113 |
| 0.8 | 0.6779 | 1.3897 | 1.2989 |
| 0.9 | 0.6218 | 1.2843 | 1.2990 |
| 1.0 | 0.5636 | 1.1543 | 1.4304 |
| 1.1 | 0.5445 | 1.1398 | 1.4592 |
| 1.2 | 0.4926 | 1.0065 | 1.4520 |
| 1.3 | 0.4594 | 0.9312 | 1.5575 |
| 1.4 | 0.4511 | 0.9230 | 1.5924 |
| 1.5 | 0.4313 | 0.8796 | 1.6252 |
| 1.6 | 0.4106 | 0.8316 | 1.6601 |
| 1.7 | 0.4014 | 0.8140 | 1.6913 |
| 1.8 | 0.3890 | 0.7866 | 1.6845 |
| 1.9 | 0.3776 | 0.7606 | 1.7450 |
| 2.0 | 0.3740 | 0.7549 | 1.7111 |

### 4.2 Setpoint Response

In this step, the load disturbance signal is considered to be zero and $G_{ff}(s)$ is determined to obtain a good setpoint response. First, for each value of $\tau_d/T$ the optimal values of $K_c$ and $T_i$ are determined using (14) and (15). Next, the setpoint response is improved using setpoint weight, $b$, which is a function of the process parameters, as shown in (16).

$$b = f_3(K_p, \tau_d, T) \qquad (16)$$

This equation can be simplified to (17), using dimensional analysis.

$$b = g_3(\frac{\tau_d}{T}) \qquad (17)$$

Using a numerical optimization technique, the optimal value of $b$ is determined so that the objective function in (9) is minimized. Table 1 and Figure 4 show the optimal values of $b$ versus $\tau_d/T$.

Using the least squares method, optimal value of $b$ can be calculated from (18).

$$b = -0.34(\tau_d/T)^2 + 1.25(\tau_d/T) + 0.53 \qquad (18)$$

There is no need to employ setpoint weighting if the setpoint response is good. The setpoint signal is not weighted if the value of $b$ is chosen equal to one. Hence, the setpoint weight will not be far from one if the setpoint signal is fairly good. However, for small values of $\tau_d/T$, the dimensionless gain, $K_pK_c$, given by (14) is large to reject load disturbance signals well. Therefore, the setpoint response is expected to be too oscillatory leading to a value of $b$ which is far from one.

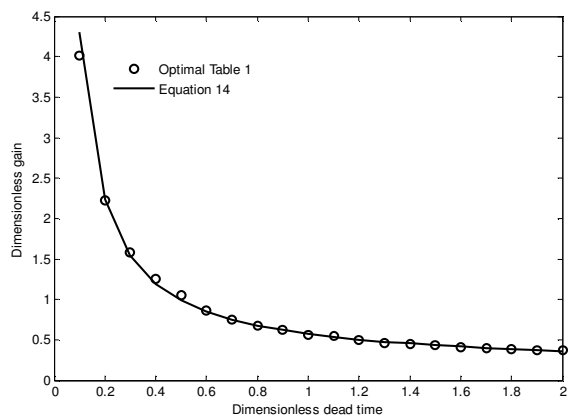

Fig. 2. Dimensionless gain versus dimensionless dead time.

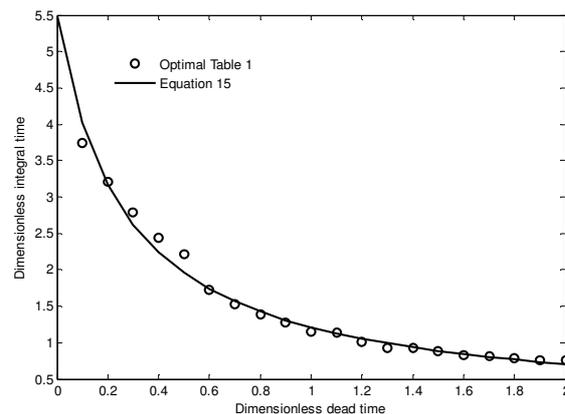

Fig. 3. Dimensionless integral time versus dimensionless dead time.

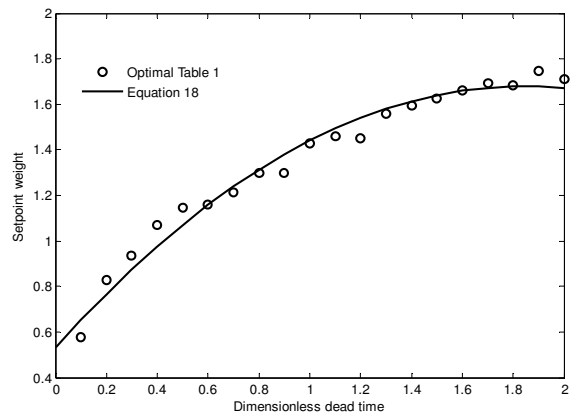

Fig. 4. Setpoint weight versus dimensionless dead time.

## 5 INTEGRATING PROCESSES

If the time constant, $T$, becomes very large, a FOPDT process is converted to an integrating process with dead time, as shown in (19).

$$G_p(s) = \lim_{T \to \infty} \frac{K_p e^{-\tau_d s}}{Ts+1} = \frac{K_p' e^{-\tau_d s}}{s} \quad (19)$$

where $K_p'$ is given by (20).

$$K_p' = \frac{K_p}{T} \quad (20)$$

Therefore, PI tuning formulae for the integrating process in (19) are obtained by using (14), (15) and (18) for a very large time constant, as shown in (21), (22) and (23).

$$K_p' K_c = \frac{0.415}{\tau_d} \quad (21)$$

$$T_i = \frac{1.48}{0.27} \tau_d \approx 5.48 \tau_d \quad (22)$$

$$b = 0.53 \quad (23)$$

## 6 SIMULATION RESULTS

In this section, performance of the proposed method is compared with that of the method presented in [9], which is one of the most prevalent techniques in PI control tuning. For simplicity, the latter method is abbreviated as APH. Both methods aim to reject load disturbance signals and improve setpoint responses through setpoint weighting whilst having a constraint on maximum sensitivity of $M_s = 1.6$.

**Example 1:** Consider the third order process $G_1(s) = \frac{1}{(s+1)^3}$. To obtain PI parameters suggested by the proposed method, the transfer function should be approximated by a FOPDT model. A simple method based on analysis of the open loop step response is given in [10]. Parameters of the FOPDT model are obtained using the following equations.

$$K_p = y_\infty \quad (24)$$
$$\tau_d = 2.8 t_1 - 1.8 t_2 \quad (25)$$
$$T = 5.5(t_2 - t_1) \quad (26)$$

where $y_\infty$ is the final value of the step response of the process and $t_1$ ($t_2$) is the time when the output attains 28% (40%) of its final value. Applying this model reduction method to $G_1(s)$, its FOPDT approximation is given by $\hat{G}_1(s) = \frac{e^{-1.039s}}{2.448s+1}$. The closed-loop step responses given by the proposed and APH methods are shown in Figure 5. The comparison results are shown in Table 2.

Tuning is a trade off between conflicting design objectives. Fast speed of response and good load disturbance rejection are design goals in conflict with good robustness [11]. As shown in Table 2 and Figure 5, the proposed controller results in a smaller value of IAE, a faster response and a better load disturbance rejection but at the cost of having a larger maximum sensitivity.

An advantage of the proposed method is that the controller parameters are directly given by (14), (15) and (18) for FOPDT processes. For a higher order process, the tuning formulae can be used after an appropriate model reduction. However, parameters of the APH controller are not explicitly given by a set of tuning formulae. They should be computed through a procedure.

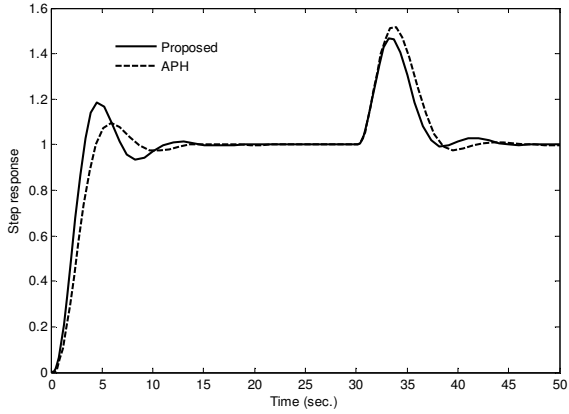

Fig. 5. Closed-loop step responses for $G_1(s)$.

TABLE 2
PERFORMANCE COMPARISION OF THE PROPOSED AND APH METHODS FOR $G_1(s)$

| Variables | Methods | |
|---|---|---|
| | Proposed | APH |
| $K_c$ | 1.210 | 0.862 |
| $T_i$ | 2.280 | 1.870 |
| $b$ | 1.090 | 0.930 |
| $M_s$ | 1.741 | 1.600 |
| IAE | 3.071 | 3.378 |

**Example 2:** Consider an integrating process with dead time $G_2(s) = \dfrac{e^{-s}}{s}$.

The closed-loop step responses resulted from the proposed and APH methods are shown in Figure 6.

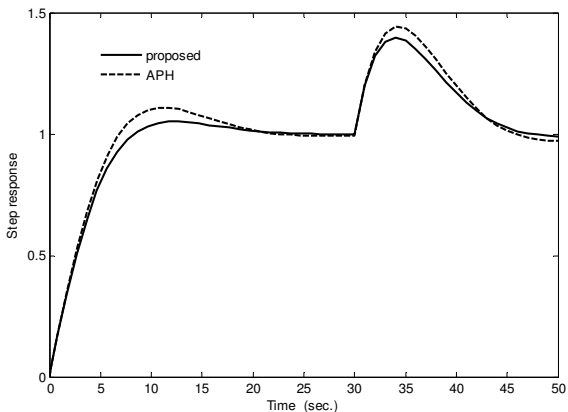

Fig. 6. Closed-loop step responses for $G_2(s)$.

The comparison results are shown in Table 3. Considering Figure 6 and Table 3, it can be concluded that the proposed controller performs better in load disturbane rejection and gives a smaller IAE, however, a slightlyfaster setpoint response and a smaller $M_s$ are given by the APH controller.

TABLE 3
PERFORMANCE COMPARISION OF THE PROPOSED AND APH METHODS FOR $G_2(s)$

| Variables | Methods | |
|---|---|---|
| | Proposed | APH |
| $K_c$ | 0.415 | 0.351 |
| $T_i$ | 5.480 | 4.967 |
| $b$ | 0.530 | 0.617 |
| $M_s$ | 1.638 | 1.590 |
| IAE | 6.934 | 8.439 |

**Example 3:** Consider a third order integrating processs $G_3(s) = \dfrac{1}{s(s+1)^2}$, where its FOPDT approximatation, given in [11], is as follows.

$$\hat{G}_3(s) = \dfrac{e^{-1.5s}}{s}$$

The closed-loop step responses resulting from the proposed and APH methods are depicted in Figure 7. The comparison results are shown in Table 4. It can be seen that the proposed controller results in a smaller value of IAE, a faster response and a better load disturbance rejection but at the expense of having a larger maximum sensitivity.

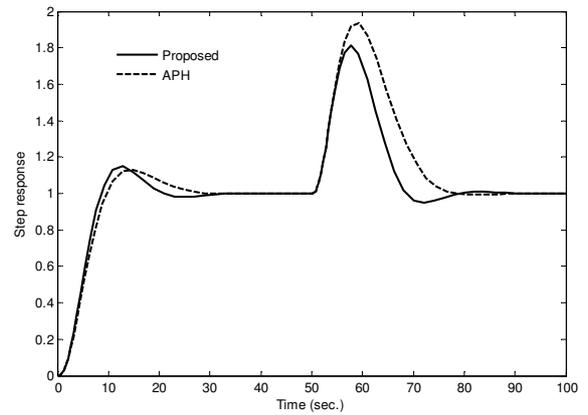

Fig. 7. Closed-loop step responses for $G_3(s)$.

TABLE 4
PERFORMANCE COMPARISION OF THE PROPOSED AND APH METHODS FOR $G_3(s)$

| Variables | Methods | |
|---|---|---|
| | Proposed | APH |
| $K_c$ | 0.277 | 0.231 |
| $T_i$ | 8.220 | 10.70 |
| $b$ | 0.530 | 0.640 |
| $M_s$ | 1.844 | 1.599 |
| IAE | 14.13 | 18.18 |

# 7 Conclusion

Control design goals can be achieved using numerical optimization methods such as particle swarm optimization. Using this technique, an efficient numerical method to obtain robust PI tuning formulae for FOPDT processes was presented in this paper. The design method was based on optimal load disturbance rejection. In order to obtain a robust controller, a constraint on the maximum sensitivity was employed. In addition, the design method could deal with setpoint response through setpoint weighting. The design procedure had two main steps. In the first step, PI controller parameters were determined such that the IAE criterion to a load disturbance step was minimized and the robustness constraint on the maximum sensitivity was satisfied. In the second step, good setpoint regulation was achieved by using a two-degree of freedom control scheme. Simulation studies for three examples showed that the proposed PI controller could effectively deal with conflicting design requirements. Although the proposed PI tuning formulae were optimal for FOPDT processes, it was also shown that they could result in good controllers for higher order processes.

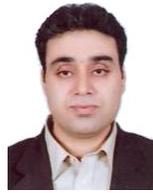

**Saeed Tavakoli** received his BSc and MSc degrees in electrical engineering from Ferdowsi University of Mashhad, Iran in 1991 and 1995, respectively. In 1995, he joined the University of Sistan and Baluchestan, Iran. He earned his PhD degree in electrical engineering from the University of Sheffield, England in 2005. As an assistant professor at the University of Sistan and Baluchestan, his research interests are space mapping optimization, multi-objective optimization, control of time delay systems, PID control design, robust control, and jet engine control. Dr. Tavakoli has served as a reviewer for several journals including IEEE Transactions on Automatic Control, IEEE Transactions on Control Systems Technology, IET Control Theory & Applications, and a number of international conferences.

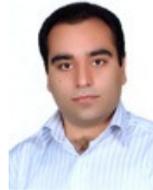

**Amir Banookh** obtained his BSc degree in electrical engineering from University of Sistan and Bluchestan, Iran in 2009. Currently, he is an MSc student at the University of Sistan and Baluchestan, Iran. His areas of research include particle swarm optimization and control design.